\documentclass[pra,superscriptaddress,reprint,floatfix]{revtex4-1}
\usepackage{bm} 
\usepackage{graphicx}

\usepackage{epstopdf}

\begin{document}

\title{Tight binding models for ultracold atoms in honeycomb optical lattices}

\author{Julen Iba\~nez-Azpiroz}
\affiliation{Depto. de F\'isica de la Materia Condensada, Universidad del Pais Vasco, UPV/EHU, 48080 Bilbao, Spain}
\affiliation{Donostia International Physics Center (DIPC), 20018 Donostia, Spain}

\author{Asier Eiguren}
\affiliation{Depto. de F\'isica de la Materia Condensada, Universidad del Pais Vasco, UPV/EHU, 48080 Bilbao, Spain}
\affiliation{Donostia International Physics Center (DIPC), 20018 Donostia, Spain}

\author{Aitor Bergara}
\affiliation{Depto. de F\'isica de la Materia Condensada, Universidad del Pais Vasco, UPV/EHU, 48080 Bilbao, Spain}
\affiliation{Donostia International Physics Center (DIPC), 20018 Donostia, Spain}
\affiliation{Centro de F\'{i}sica de Materiales CFM, Centro Mixto CSIC-UPV/EHU, 20018 Donostia, Spain}

\author{Giulio Pettini}
\affiliation{Dipartimento di Fisica e Astronomia, Universit\`a di Firenze,
and INFN, 50019 Sesto Fiorentino, Italy}

\author{Michele Modugno}
\affiliation{Depto. de F\'isica Te\'orica e Historia de la Ciencia, Universidad del Pais Vasco, UPV/EHU, 48080 Bilbao, Spain}
\affiliation{IKERBASQUE, Basque Foundation for Science, 48011 Bilbao, Spain}

\begin{abstract}
We discuss how to construct tight-binding models for ultra cold atoms in honeycomb potentials, by means of the maximally localized Wannier functions (MLWFs) for composite bands introduced by \citet{marzari1997}. In particular, we work out the model with up to third-nearest neighbors, and provide explicit calculations of the MLWFs and of the tunneling coefficients for the graphene-lyke potential with two degenerate minima per unit cell. Finally, we discuss the degree of accuracy in reproducing the exact Bloch spectrum of different tight-binding approximations, in a range of typical experimental parameters.
\end{abstract}

\date{\today}

\maketitle

\textit{Introduction.}
Ultracold atoms in optical lattices are routinely employed as simulator of condensed matter physics, thanks to the possibility of engineering several geometry configurations and of tuning the system parameters with great flexibility and precision \cite{bloch2008,lewenstein2007}. In particular, honeycomb lattices are attracting an increasing interest as they allow to mimic the physics of graphene, where the presence of topological defects in momentum space, the so-called Dirac points, leads to remarkable relativistic effects \cite{wunsch2008,lee2009,soltan-panahi2011,soltan-panahi2011b,gail2012,tarruell2012,lim2012}.

Despite the fact that the potentials describing optical lattices are continuous and can be expressed in simple analytic forms as the combination of a number of sinusoidal potentials, from the theoretical point of view it is often convenient to describe the system by means of a tight binding approach on a discrete lattice.  In fact, the potential intensity can be tuned to sufficiently high values so to localize the atoms in the lowest vibrational states of the potential wells, justifying a description in terms of tunneling coefficients related to the hopping between neighboring sites (the potential minima), and interaction strengths which characterize the onsite interaction among the atoms \cite{bloch2008}. This applies also for the case of honeycomb lattices, where a number of tight binding approaches have been considered recently \cite{lee2009,gail2012,lim2012,hasegawa2012}, in analogy to the case of graphene \cite{wallace1947,cloizeaux1963,cloizeaux1964,reich2002}.

A crucial ingredient for the connection between the continuos and discrete versions of the system hamiltonian is the existence of a basis of functions localized around the potential minima.
This is important not only conceptually - in order to justify the tight binding expansion - 
but also from the practical point of view, as a precise knowledge of the basis functions is needed to connect the tight binding coefficients with the actual experimental parameters \cite{modugno2012}. In the case of optical lattices with a cubic-like arrangement (with a single well per unit cell) this basis is provided by the exponentially decaying Wannier functions discussed by Kohn \cite{kohn1959,he2001,bloch2008}, from which one can derive analytic expressions for the tight-binding coefficients \cite{gerbier2005}. 
However, in general this approach may fail when the potential has more than one well per unit cell. For example, for the case of a honeycomb potential with two degenerate minima in the unit cell, the Kohn-Wannier cannot be associated to a single lattice site as they occupy both cells for symmetry reasons \cite{cloizeaux1963,cloizeaux1964}.

A powerful approach, that is widely used for describing real 
material structures, is represented by the maximally localized Wannier functions (MLWFs) 
introduced by Marzari and Vanderbilt \cite{marzari1997}. 
The MLWFs are obtained by minimizing the spread of a set of 
generalized Wannier functions by means of a 
suitable gauge transformation of the Bloch eigenfunctions for a composite band. 
Due to their exponential decay \cite{brouder2007,panati2011},
the MLWFs provide an optimal basis set
for tight-binding models and is widely employed in condensed matter physics \cite{marzari2012}.

In this paper we discuss the tight-binding expansion up to third-nearest neighbors for ultracold atoms in honeycomb lattices, and explicitly calculate the MLWFs and the tunneling coefficients for a potential with two degenerate minima in the unit cell, by using the WANNIER90 package \cite{mostofi2008}. For the tunneling coefficients we also provide an analytic expression in terms of the lattice intensity, obtained from a fit of the data. Then we discuss the validity of different tight-binding approximations - including only the nearest neighbor tunneling or up to the third-nearest neighbor -  in terms of the experimental parameters.

\textit{Tight binding approach for the honeycomb potential.}
\begin{figure}[t]
\centerline{\includegraphics[width=0.7\columnwidth]{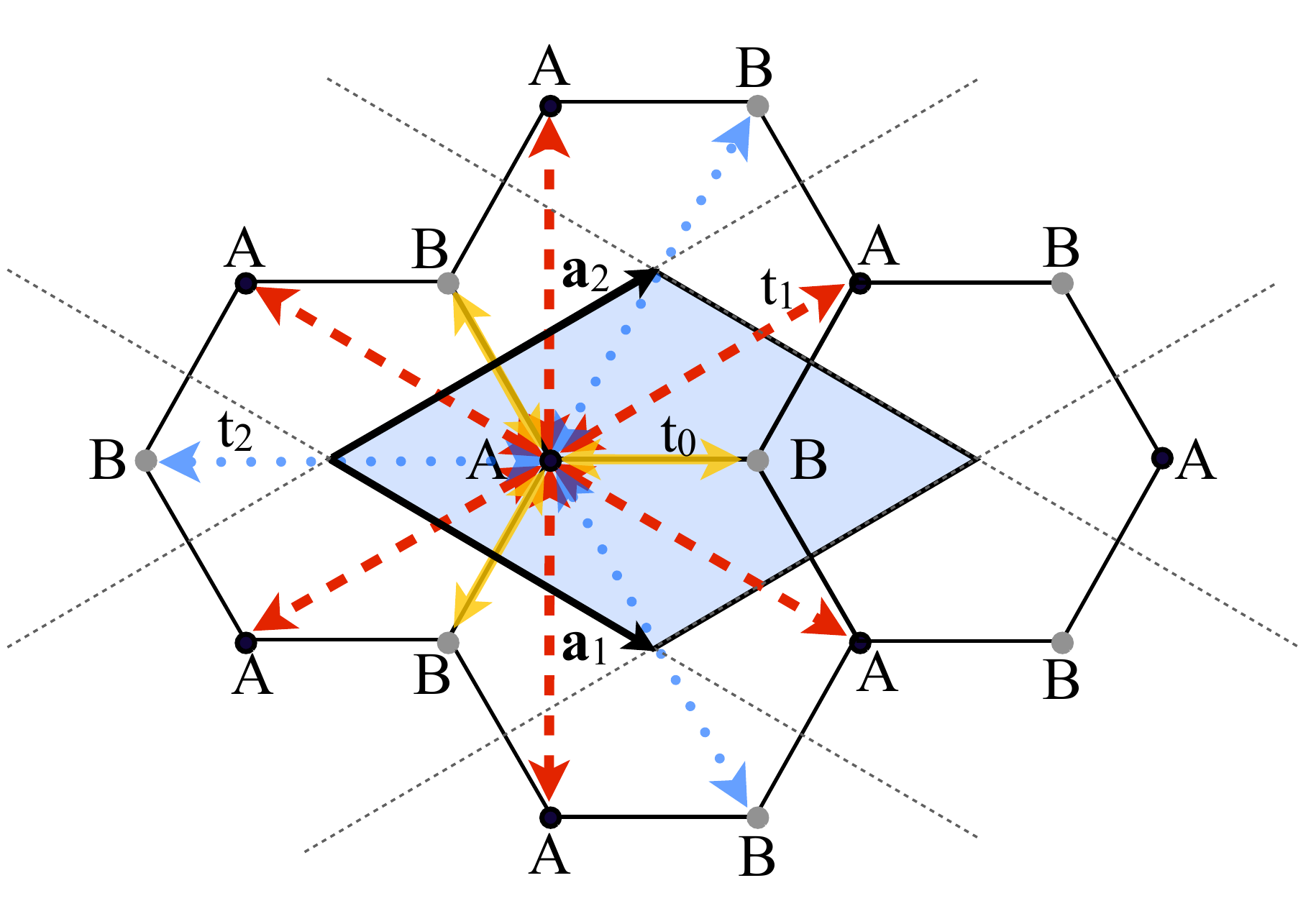}}
\caption{(Color online) Sketch of the honeycomb lattice structure and the diamond-shaped elementary cell with basis $A$ and $B$. The length of each side of the hexagon is  
$a=4\pi/(3\sqrt{3}k_{L})$. The different tunnelings defined in the text are indicated for a site of type A.}
\label{fig:honeycomb}
\end{figure}
Let us start by considering the two-dimensional graphene-like lattice discussed by \citet{lee2009}
\begin{equation}
V(\bm{r})=3
+2\cos\left[\left(\bm{b}_{1}+\bm{b}_{2}\right)\cdot\bm{r}\right]
+2\sum_{i=1,2}\cos\left(\bm{b}_{i}\cdot\bm{r}\right)
\label{eq:pot}
\end{equation}
where $\bm{r}=(x,y)$ and $\bm{b}_{1/2}=\sqrt{3}k_{L}({\bm{e}_{x}\mp\sqrt{3}{\bm{e}}_{y}})/{2}$ ($k_{L}$ being the modulus of the laser wavevectors),
corresponding to a honeycomb structure, with a diamond-shaped elementary cell with basis $A$ and $B$ 
as shown in Fig. \ref{fig:honeycomb}. The Bravais lattice $\cal{B}$ in real space is generated by two fundamental
 vectors ${\bf{a}}_{1},{\bf{a}}_{2}$ defined by ${\bf{a}}_{i}\cdot{\bf{b}}_{j}=2\pi\delta_{ij}$, so that
${\cal{B}}=\{j_{1}{\bf{a}}_{1}+j_{2}{\bf{a}}_{2}\Big| j_{1},j_{2}=0,\pm 1,\pm 2 \dots\}$ \cite{lee2009}.

The starting point for constructing a tight-binding model is the many-body hamiltonian for bosonic or fermionic particles, described by the field operator $\hat{\psi}(\bm{r})$. In the following we will focus on the single-particle term
\begin{equation}
\hat{\cal{H}}_{0}=\int d\bm{r}~{\hat{\psi}}^\dagger(\bm{r})\hat{H}_{0}{\hat{\psi}}(\bm{r})
\label{eq:genham}
\end{equation}
as the mapping onto the tight-binding model is determined by the spectrum of the single particle hamiltonian $H_{0}=-(\hbar^{2}/2m)\nabla^{2} + sE_{R}V(\bm{r})$ \cite{jaksch1998}.
Here $s$ represents the potential amplitude in units of the \textit{recoil energy} $E_{R}=\hbar^{2}k_{L}^{2}/2m$.

In general, when the potential wells are deep enough, the hamiltonian (\ref{eq:genham}) can be conveniently mapped onto a \textit{tight-binding model} defined on the discrete lattice corresponding to the potential minima, by expanding the field operator in terms of a set of functions $\{w_{\bm{j}\nu}(\bm{r})\}$ localized at each minimum, as
\begin{equation}
\hat{\psi}(\bm{r})\equiv \sum_{\bm{j}\nu}{\hat{a}}_{\bm{j}\nu}w_{\bm{j}\nu}(\bm{r})
\label{eq:psiexpf}
\end{equation}
where ${\bm{j}}=(j_1,j_2)$ labels the cell and $\nu$ is a band index.
In Eq. (\ref{eq:psiexpf}) $\hat{a}_{\bm{j}\nu}^{\dagger}$ ($\hat{a}_{\bm{j}\nu}$) represent the creation (destruction) operators of a single particle in the cell $\bm{j}$, and satisfy the usual commutation rules $[\hat{a}_{\bm{j}\nu},\hat{a}^{\dagger}_{\bm{j}'\nu'}]=\delta_{\bm{j}\bm{j}'}\delta_{\nu\nu'}$ (following from those for the field $\hat{\psi}$).

In order to construct a basis of localized functions at each site of the  lattice
here we consider the MLWFs for a composite band discussed by \citet{marzari1997}. 
These are a set of generalized Wannier functions $w_{\bm{j}\nu}$,  
defined from a linear combination of Bloch eigenstates $\psi_{n\bm{k}}$, namely \cite{marzari1997,modugno2012}
\begin{eqnarray}
w_{\bm{j}\nu}(\bm{r})&=&\frac{1}{\sqrt{S_{\cal B}}}
\int_{S_{\cal B}} \!\!d\bm{k} ~e^{-i\bm{k}\bm{R}_{\bm{j}}}\sum_{m=1}^{N}U_{\nu m}(\bm{k})\psi_{m\bm{k}}(\bm{r})
\label{eq:mlwfs}
\end{eqnarray}
with $S_{\cal B}$ indicating the first Brillouin zone, and $U\in U(N)$ being a gauge transformation
that minimizes the Marzari-Vanderbilt localization functional 
$\Omega=\sum_{\nu}\left[\langle \bm{r}^2\rangle_{\nu}-\langle \bm{r}\rangle_{\nu}^{2}\right]$ \cite{marzari1997}.
The real character and the exponential decay of MLWF functions has been demonstrated 
under very general assumptions \cite{brouder2007,panati2011}. Thus, these functions represent an optimal basis for tight-binding models. The real character of the calculated Wannier functions has been confirmed in our computations.

\begin{figure}[b!]
\centerline{\includegraphics[width=0.65\columnwidth]{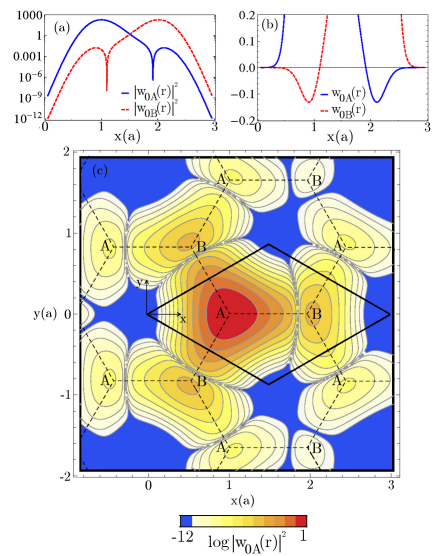}}
\caption{(Color online) Example of the calculated MLWFs for $s=15$. (a) 
Profile of the MWLWs $|w_{\bm{0}A}(\bm{r})|^{2}$ (solid, blue) and 
$|w_{\bm{0}B}(\bm{r})|^{2}$ (dashed, red) along the line joining the $A$
and $B$ sites ($y=0$) in the original unit cell. 
(b) Profile of $w_{\bm{0}A}(\bm{r})$ and $w_{\bm{0}B}(\bm{r})$ along the same path as in (a) 
with a zoom into the small values of the
MLWFs. Note that $w_{\bm{0}A}(\bm{r})$ ($w_{\bm{0}B}(\bm{r})$) becomes 
negative in the neighborhood of site $B$ ($A$).
(c) Contour plot of the function $\log|w_{\bm{0}A}(\bm{r})|^{2}$.
The solid and dashed lines depict the original unit cell and
honeycomb lattice, respectively.
See text for explanations. }
\label{fig:mlwfs}
\end{figure}

In our case, since the honeycomb unit cell contains two potential minima, $A$ and $B$, 
we can construct a basis of MLWFs by considering a composite band consisting 
of the two lowest Bloch bands, that is $N=2$. 
The mentioned Bloch sub-bands have been obtained with a 
modified version of the QUANTUM-ESPRESSO package \cite{espresso}
intended to solve the single particle Schr\"{o}dinger
equation associated to the external potential (\ref{eq:pot}).
We consider a plane wave expansion of the Bloch states, reaching convergence with an energy cutoff corresponding to 
$E_c$=10.5 $E_{R}$ and a {\bf k}-point mesh of 14$\times$14.
As a next step, the MLWFs have been computed considering the
WANNIER90 program ~\cite{mostofi2008,marzari2012}.
The typical shape of the calculated MLWFs is 
shown in Fig. \ref{fig:mlwfs}.

The strong localization of $|w_{\bm{0}\nu}(\bm{r})|^{2}$ ($\nu=A,B$)
around sites $A$ and $B$, and their exponential decay are clearly visible in panel (a).
Panel (c) shows the distribution
of $|w_{\bm{0}A}(\bm{r})|^{2}$ around 
the original unit cell $\bm{j}=\bm{0}$;
the figure reveals an appreciable overlap of the MLWF
with the neighboring $B$ and $A$ sites, indicated respectively
by the yellow and red arrows in Fig. \ref{fig:honeycomb}.
In addition, the MLWFs are characterized by nodes in passing from sites of type A to B, corresponding to a change of sign (see panel (b)).

The approach followed here of including two Bloch bands 
is the minimal approximation, and corresponds to the generalization of the
\textit{single band approximation} usually employed for cubic-type lattices \cite{bloch2008}.
Within this approximation,
the hamiltonian (\ref{eq:genham}) can be written as 
\begin{equation}
\hat{\cal{H}}_{0} \simeq \sum_{\nu\nu'=A,B}\sum_{\bm{j,j'}}\hat{a}^{\dagger}_{\bm{j}\nu}\hat{a}_{\bm{j}'\nu'}
\langle w_{\bm{j}\nu}|\hat{H}_{0}|w_{\bm{j'}\nu '}\rangle\equiv\hat{\cal{H}}_{0}^{tb}
\label{eq:htb}
\end{equation}
where the expansion coefficients in the above equation correspond 
to the onsite energies 
$E_{\nu}=\langle w_{\bm{j}\nu}|\hat{H}_{0}|w_{\bm{j}\nu}\rangle$,  and to the tunneling amplitudes between different sites. Both quantities are real thanks to the properties of the MLWFs.
Here we will include up to the third-nearest neighbor tunnelings $T_{\nu\nu'}^{\bm{i}}\equiv-\langle w_{\bm{j}\nu}|\hat{H}_{0}|w_{\bm{(j+i)}\nu'}\rangle$, with
${\bm{i}}\equiv(0,\pm 1;0, \pm 1)$ (they depend only on the relative distance owing to the uniformity of the lattice). Then, the tunnelings can be divided in three classes, as shown in Fig. \ref{fig:honeycomb} for the case $\nu=A$.

\textit{(i)} 
Terms between $A(B)$ and the three nearest neighbors of type $B(A)$, yellow arrows in Fig. \ref{fig:honeycomb}, \textit{e.g.} \begin{eqnarray}
 t_{0}&=&-\langle w_{\bm{j}A}|\hat{H}_{0}|w_{\bm{j}B}\rangle
\end{eqnarray}

\textit{(ii)}
Terms $t_{\nu}$ between sites of the same type ($A$ or $B$) within neighboring cells, red arrows in Fig. \ref{fig:honeycomb}, \textit{e.g.}
\begin{eqnarray}
t_{\nu}&=&-\langle w_{(\bm{j}_{1}+\bm{1},\bm{j}_{2}){\nu}}|\hat{H}_{0}|w_{(\bm{j}_{1},\bm{j}_{2}){\nu}}\rangle
\end{eqnarray}
In general the two tunneling coefficients $t_{A}$ and $t_{B}$ are different. As a specific example, here we will explicitly compute them for the case of degenerate minima where $t_{A}=t_{B}$. In this case we can set $t_{1}\equiv -t_{\nu}$, where the sign is chosen in order to have $t_{1}$ positive defined (see Fig. \ref{fig:mlwfs} and the discussion about the sign of the MLWFs).

\textit{(iii)}
Terms connecting $A(B)$ to $B(A)$ at opposite corners of the hexagon, blue arrows in Fig. \ref{fig:honeycomb}, \textit{e.g.}
\begin{eqnarray}
 t_{2}&=&-\langle w_{(\bm{j}_{1},\bm{j}_{2})A}|\hat{H}_{0}|w_{(\bm{j}_{1}-\bm{1},\bm{j}_{2}-\bm{1})B}\rangle.
\end{eqnarray}

We remark that the above derivation of the tight-binding model is valid in general for any potential with a honeycomb structure, with two minima per unit cell, and not just for the potential with two degenerate minima in Eq. (\ref{eq:pot}). In the following we will consider explicitly the latter case in order to provide a specific example. 

\begin{figure}[t]
\centerline{\includegraphics[width=0.7\columnwidth]{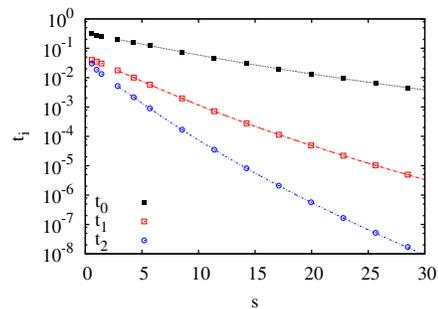}}
\caption{(Color online) Behavior of the various tunnelings as a function of the lattice intensity $s$. The lines are the result of a fit of the numerical data, and those for $t_{0}$ and $t_{1}$ coincide with that extracted from  a fit of the Bloch spectrum (see text).}
\label{fig:tunnel}
\end{figure}

The behavior of the different tunneling coefficients as a function of the lattice intensity $s$ is shown in Fig. \ref{fig:tunnel}.
In order to extract from the numerical values an analytic formula we consider a fit
of the type $t_{i}= A s^{\alpha} e^{-\beta\sqrt{s}}$ ($i=0,1,2$),
in the range $s>3$, with $A$, $\alpha$, and $\beta$ as fitting parameters.
For $t_{0}$ we find 
\begin{equation}
t_{0}= 1.16 s^{0.95} e^{-1.634\sqrt{s}},
\label{eq:t0}
\end{equation}
that has to be compared with the semiclassical estimate
of \citet{lee2009},  $|t_{0}|= 1.861 s^{0.75} e^{-1.582\sqrt{s}}$ (see Eq. (38) in \cite{lee2009}).
In the range of $s$ considered here, we find that the latter overestimates the actual value in (\ref{eq:t0}) by about $8\%$ for $s=30$, up to about $40\%$ for $s=3$.
For the other two terms we get
\begin{eqnarray}
t_{1} &=& 0.78 s^{1.85} e^{-3.404\sqrt{s}},\\
t_{2} &=& 1.81 s^{2.75} e^{-5.196\sqrt{s}}.
\end{eqnarray}
These three fit are shown as dashed lines in Fig. \ref{fig:tunnel}.

\begin{figure}[t!]
\centerline{\includegraphics[width=0.9\columnwidth]{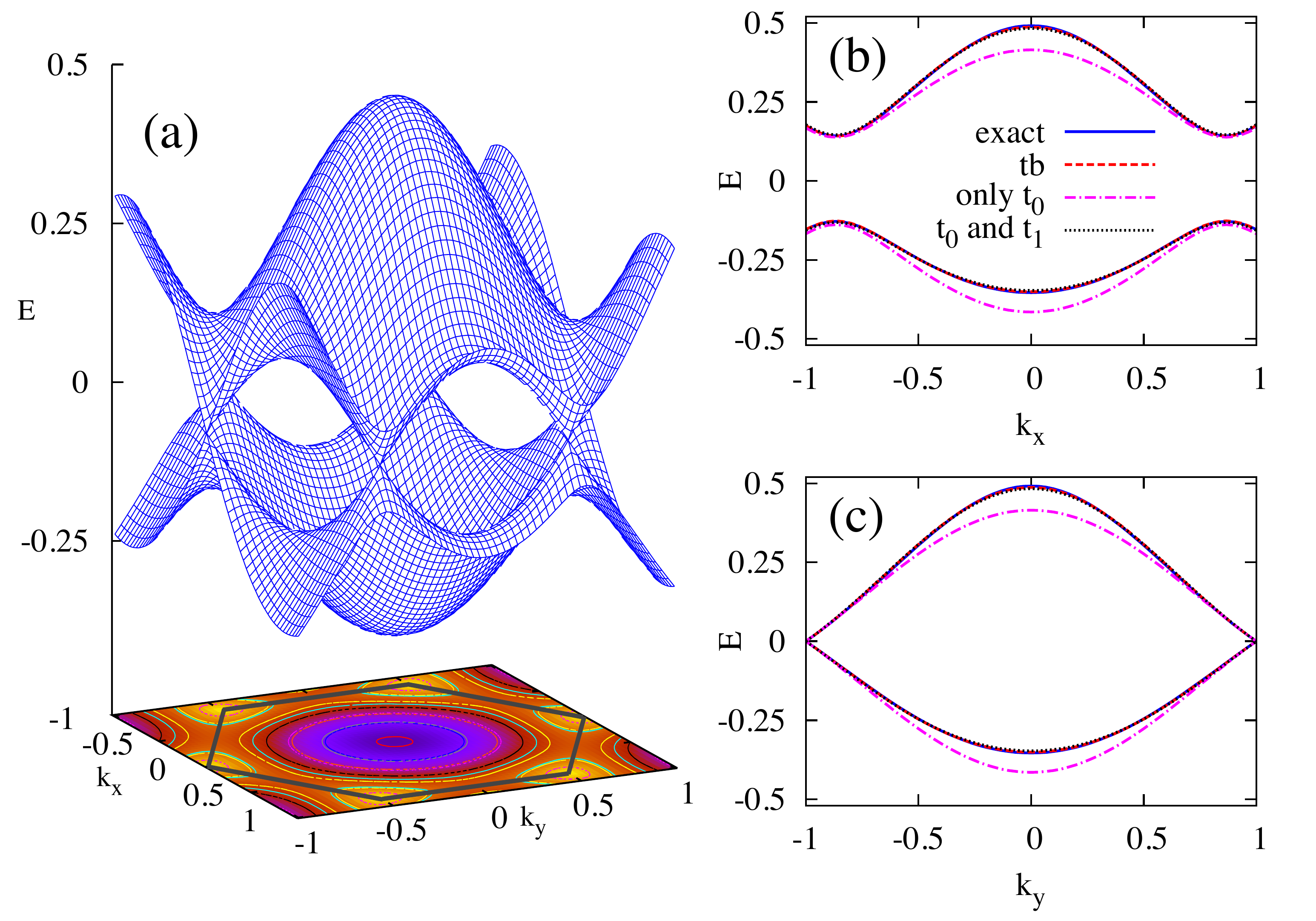}}
\caption{(Color online) (a) Bloch spectrum $E_{\pm}(\bm{k})$ of the lowest two bands for $s=5$. The hexagon represents the first Brillouin zone. (b,c) Cut for $k_{y}=0$ and $k_{x}=0$, respectively (blue, solid  lines). The latter are compared with the prediction of the full tight-binding model (red, dashed  line), that with $t_{0}$ and $t_{1}$ (black, dotted), and with just $t_{0}$ (magenta, dot-dashed). Note the asymmetry of the two bands.}
\label{fig:bands}
\end{figure}

\textit{Tight-binding spectrum.}
A convenient way to check the regime of validity of a given tight binding approximation is to compare its prediction for the energy spectrum with the exact Bloch spectrum.
The latter can be readily computed by means of a standard Fourier decomposition \cite{lee2009}.
The typical structure of the two lowest bands $E_{\pm}(\bm{k})$ is shown in Fig. \ref{fig:bands}.
It is characterized by Dirac points at the vertices of the Brillouin zone (a regular hexagon), where the local dispersion is linear and the two bands are degenerate \cite{lee2009}. For convenience, we fix $E_{\pm}(\bm{k}_{D})=0$ at the Dirac points.

The tight-binding spectrum can be derived as follows \cite{modugno2012}. By defining 
$\hat{b}_{\nu\bm{k}}=(\sqrt{S_{\cal B}}/2\pi)
\sum_{\bm{j}} ~e^{-i\bm{k}\cdot\bm{R}_{\bm{j}}}\hat{a}_{\bm{j}\nu}
$ the hamiltonian ${\hat{\cal{H}}}_0^{tb}$ in Eq. (\ref{eq:htb}) can be written as
\begin{equation}
\hat{\cal{H}}_{0}^{tb}=\sum_{\nu\nu'=A,B}\int_{S_{\cal B}} d\bm{k}~h_{\nu\nu'}(\bm{k})
\hat{b}_{\nu\bm{k}}^{\dagger}\hat{b}_{\nu'\bm{k}}
\end{equation}
with
$h_{\nu\nu'}(\bm{k})=-\sum_{\bm{i}}e^{i\bm{k}\cdot\bm{R}_{\bm{i}}}T_{\nu\nu'}^{\bm{i}}$.
Finally, the matrix $h_{\nu\nu'}(\bm{k})$ turns out of the form \cite{lee2009}
\begin{equation}
h_{\nu\nu'}(\bm{k})=\left(\begin{array}{cc}
 \epsilon_{A}(\bm{k}) & z(\bm{k}) \\
 z^{*}(\bm{k}) & \epsilon_{B}(\bm{k})
\end{array}\right).
\label{eq:hmatrix}
\end{equation}

As for the diagonal terms, the leading contribution comes from the onsite energies
$E_{\nu}$,
that can be conveniently written as $E_{A/B}=\pm\epsilon$, by shifting the total energy by an overall constant. 
In addition, there is a correction due to $t_{\nu}$,
 so that eventually we get
\begin{equation}
\epsilon_{A/B}(\bm{k})=E_{A/B}-t_{A/B} F(\bm{k}),
\end{equation}
\begin{equation}
F(\bm{k})=2 \cos\left(\bm{k}\cdot(\bm{a}_{1}-\bm{a}_{2})\right) + 2\sum_{i=1,2}\cos\left(\bm{k}\cdot\bm{a}_{i}\right).
\end{equation}

Instead, the off-diagonal terms get the leading contribution from the term proportional to $t_{0}$ \cite{lee2009}, and a correction due to $t_{2}$, $z(\bm{k})=t_{0}Z_{0}(\bm{k}) +t_{2}Z_{2}(\bm{k})$, with
\begin{eqnarray}
Z_{0}(\bm{k})&=&1+e^{-i\bm{k}\cdot\bm{a}_{1}}+e^{-i\bm{k}\cdot\bm{a}_{2}}\nonumber\\
Z_{2}(\bm{k})&=&e^{-i\bm{k}\cdot(\bm{a}_{1}+\bm{a}_{2})}+e^{-i\bm{k}\cdot(\bm{a}_{1}-\bm{a}_{2})}+e^{-i\bm{k}\cdot(\bm{a}_{2}-\bm{a}_{1})}.\nonumber
\end{eqnarray}

Finally, by diagonalizing the matrix $h_{\nu\nu'}(\bm{k})$ and defining 
$t_{\pm}\equiv({t_{A}\pm t_{B}})/{2}$, we get the following expression for the tight-binding spectrum
\begin{equation}
\epsilon_{\pm}(\bm{k})=-t_{+}F(\bm{k})\pm\sqrt{\displaystyle{(t_{-}F(\bm{k})-\epsilon)^{2}+|z(\bm{k})|^{2}}}.
\end{equation}

Again, this is valid for a generic honeycomb structure with two minima per unit cell. For the particular case of degenerate minima, as for the potential in (\ref{eq:pot}), this expression further reduces to 
\begin{equation}
\bar{\epsilon}_{\pm}(\bm{k})=t_{1}F(\bm{k})\pm |t_{0}Z_{0}(\bm{k}) +t_{2}Z_{2}(\bm{k})| + 3t_{1}
\label{eq:enerdeg}
\end{equation}
where the last term has been added in order to make the energy vanishing at the Dirac points, consistently with the definition of the Bloch spectrum. A specific example is shown in  Fig. \ref{fig:bands}(b,c), for $s=5$. The figure shows that the tight-binding model with just $t_{0}$ is not sufficient to reproduce the band structure, and that at least the inclusion of $t_{1}$ is needed. In particular, the latter is necessary to account for the band asymmetry, see Eq. (\ref{eq:enerdeg}).
It also remarkable to notice that the values of $t_{0} + t_{2}$ and $t_{1}$ can be extracted from a fit  of the Bloch spectrum at $\bf k=0$, by using the expression (\ref{eq:enerdeg}). In fact, we have
$t_{0}+t_{2}=(\bar{\epsilon}_{+}(\bm{0})-\bar{\epsilon}_{-}(\bm{0}))/6$ and $t_{1}=(\bar{\epsilon}_{+}(\bm{0})+\bar{\epsilon}_{-}(\bm{0}))/18$. In addition, since $t_{2}$ is negligible with respect to $t_{0}$ (see Fig. \ref{fig:tunnel}), practically the former expression can be used as an estimate of $t_{0}$. Notably, these estimates coincide with the dashed lines shown in Fig. \ref{fig:tunnel}, therefore providing an independent check of the tunnelings calculated by means of the MLWFs. 

\begin{figure}[h!]
\centerline{\includegraphics[width=0.75\columnwidth]{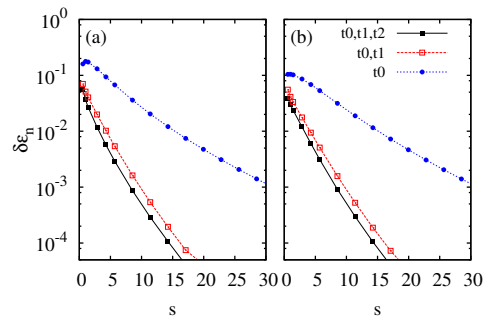}}
\caption{(Color online) Energy mismatch $\delta\varepsilon_{n}$ for the first (a) and second (b) band,
for different levels of approximation of the tight-binding model.
}
\label{fig:de}
\end{figure}

In general, one can evaluate the degree of accuracy in reproducing the exact Bloch spectrum by defining an energy mismatch as follows
\begin{equation}
\delta\varepsilon_{n}\equiv\frac{1}{\Delta E_{n}}\sqrt{\frac{1}{S_{\cal B}}\int_{S_{\cal B}}d\bm{k}\left[E_{n}(\bm{k}) -\epsilon_{n}(\bm{k})\right]^{2}}
\end{equation}
with $\Delta E_{n}$ being the $n-$th bandwidth ($n=1,2$).  
The results are shown in Fig. \ref{fig:de}. This figure shows that the tight-binding model with up to  third-nearest neighbors accurately reproduces the band structure for $s\gtrsim 3$, with an error below $1\%$. In fact, this is a range of lattice intensities where one would expect the MLWFs to localize strongly around each minimum (that is, a proper tight-binding regime). 
Then, while the inclusion of $t_{2}$ provides only a minor correction, the model with just the nearest neighbor tunneling $t_{0}$ \cite{lee2009} is clearly less accurate, reaching the level $\delta\varepsilon_{n}\lesssim1\%$ only for $s\gtrsim15$.  This may be particularly relevant for the range of parameters of current experiments \cite{tarruell2012}. 

\textit{Conclusions.}
In this article we have demonstrated the power of the maximally localized Wannier functions for composite bands in determining the parameters of tight-binding hamiltonians describing ultracold atoms in optical lattices.
The application to a honeycomb structure, directly connected to the graphene physics,
allows us to accurately parametrize the optimal tight-binding parameters, providing a thorough analysis of the range of validity of different approaches.

\textit{Acknowledgments.}
 This work has been supported by the UPV/EHU under programs 
UFI 11/55 and IT-366-07, the Spanish Ministry of
Science and Innovation through the Grant No. FIS2010-19609-C02-00
and the Basque Government through the Grant No. IT-472-10.

\bibliography{biblio}

\end{document}